\documentclass[12pt]{article} 
\usepackage[utf8]{inputenc} 
\usepackage{geometry}
\usepackage{graphicx} 
\usepackage{amssymb}
\usepackage{amsmath}
\usepackage{cite}
\usepackage{mathrsfs}
\usepackage{booktabs} 
\usepackage{array} 
\usepackage{paralist}
\usepackage{verbatim} 
\usepackage{subfig}
\usepackage{fancyhdr} 
\usepackage{sectsty}
\allsectionsfont{\sffamily\mdseries\upshape} 
\usepackage[nottoc,notlof,notlot]{tocbibind} 
\usepackage[titles,subfigure]{tocloft} 

\title{Exact solution to the Dirichlet problem for degenerating on the boundary elliptic equation of Tricomi -Keldysh type in the half-space}
\author{\bf Oleg D. Algazin}
\date{Bauman Moscow State Technical University} 

\begin{document}
\maketitle
\thispagestyle{empty}

\begin{abstract}
In the paper by means of Fourier transform method and   similarity method we solve the Dirichlet problem for a multidimensional equation wich is a generalization of the Tricomi, Gellerstedt and Keldysh equations  in the half-space, in  which   equation have elliptic type,  with the boundary condition on the boundary hyperplane where equation degenerates.The solution is presented in the form of an integral with a simple kernel which  is an approximation to the identity and self-similar solution of  Tricomi-Keldysh type equation . In particular, this formula contains a Poisson's formula, which gives the solution of the Dirichlet problem for the Laplace equation for the half-space. If the given boundary value is a generalized function of slow growth, the solution of the Dirichlet problem can be written as a convolution of this function with the kernel (if a convolution exists).
\end{abstract}


\section*{Introduction}
         In the paper is considered the multidimensional elliptic equation in the half-space
\[
y^m\Delta_xu+u_{yy}=0,~~ y>0,~~ m>-2,\eqno{(\textup{T})}
\]
where $x=(x_1,x_2,\dots,x_n)\in\mathbb{R}^n, u=u(x,y)$~  is a function of variabels~$(x,y)\in\mathbb{R}^{n+1}$, 
\[
\Delta_x=\frac{\partial^2 }{\partial x_1^2}+\dots+\frac{\partial^2 }{\partial x_n^2}
\]
is the Laplace operator on the variable $x$.
\begin{enumerate}
\item 
If $n=1, m=1$ we obtain the Tricomi equation
\[
yu_{xx}+u_{yy}=0.
\]
\item 
If $n=1, m>0$ we obtain the Gellerstedt equation
\[
y^mu_{xx}+u_{yy}=0,~~m>0.
\]
\item 
If $n=1, m<0$ the equation (T)  can be written as
\[
u_{xx}+y^{-m}u_{yy}=0,~~0<-m<2,
\]
it is a special case of the Keldych equation \cite{Kel}.

These equations are used in transonic gas dynamics \cite{Ber}, and in mathematical models of cold plasma \cite{Otw}.
\item
If $m=0$ we obtain the Laplace equation
\[
\Delta u(x,y)=0.
\]
\end{enumerate}
A bounded (as $y\to\infty$) solution of the Dirichlet problem for the Laplace equation for the half-space
\[
\Delta u(x,y)=0,~~x\in\mathbb{R}^n,~~y>0,
\]
\[
u(x,0)=\psi(x),~~x\in\mathbb{R}^n,
\]
is given by the Poisson integral \cite{Bit},\cite{Ste}
\[
u(x,y)=\frac{\Gamma((n+1)/2)}{\pi^{(n+1)/2}}\int_{\mathbb{R}^n}\frac{y\psi(t)}{(|x-t|^2+y^2)^{(n+1)/2}}dt.
\]
   A similar formula is derived by us in this paper for the solution of the Dirichlet problem for an Tricomi-Keldysh type equation (T) by means of the Fourier transform to the variables in the boundary hyperplane $y=0$ in the case $m=1$, and by the similarity method in the case of $m>-2$. In this formula, in particular, is contained the Poisson's integral formula ($m=0$) , which can also be obtained using the Fourier transform. For the case $ -2 <m <0$ this formula was earlier obtained by L.S.Parasyuk  by Fourier transform \cite{Par}. In the case of $m> 0$ in the calculation of multidimensional Fourier transformations there are great difficulties (except  the case $m = 1$, wich we consider in section 2). Therefore, we apply the similarity method. With it, in section 3, we find a self-similar solution of the equation of Tricomi -Keldysh type for any $m> -2$, which is an approximation to the identity in the space of integrable functions. Solution of the Dirichlet problem is represented as a convolution of the self-similar solution of equation of Tricomi-Keldysh type with boundary function (if thise convolution exists). The general properties of the approximation to the identity implies that in the case of bouded piecewise continuous boundary function  this convolution is written in the form of an integral and gives the classical solution of the Dirichlet problem, i.e. the boundary values of integral coincide with the boundary function at all points of continuity. In the case of the boundary function, which is a generalized function of slow growth, the convolution gives a generalized solution of the Dirichlet problem, i.e. weakly converges in the space of generalized functions of slow growth to a given boundary generalized function. In particular, the kernel is a solution of the Dirichlet problem, where the boundary function is the Dirac delta function.

      We note that  by the  similarity method  was obtained  the fundamental solutions for the Tricomi operator  ($m = 1$) in the works of J.Barros-Neto and I.M. Gelfand \cite{Gel1},\cite{Gel2},\cite{Gel3} and by Fourier transform method ($m = 1$) in the work of J.Barros-Neto and F.Cardoso \cite{Car}.

    Earlier, using the Fourier transform method we solved in our joint works the Dirichlet and Dirichlet-Neumann problem for the Laplace and Poisson equations in a multidimensional infinite layer \cite{Alg1},\cite{Alg2}.

\section{Notations and statement of the problem}
We introduce the following notations:
\[
x=(x_1+\dots+x_n)\in\mathbb{R}^n,~~(x,y)=(x_1+\dots+x_n,y)\in\mathbb{R}^{n+1},~~y\in\mathbb{R},
\]
\[
|x|=\sqrt{x_1^2+\dots+x_n^2},~~xt=x_1t_1+\dots+x_nt_n,~~dx=dx_1\dots dx_n,
\]
\[
F(t)=\mathscr{F}[f](t)=\int_{\mathbb{R}^n}f(x)e^{ixt} dx-
\]
Fourier transform of an integrable function $f(x)$ . If integrable in $x$ function $f(x,y)$ depends on the variables $x$ and $y$, then its Fourier transform with respect to $x$ will be denoted
\[
\mathscr{F}_x[f](t,y)=\int_{\mathbb{R}^n}f(x,y)e^{ixt} dx.
\]
Similarly we define the inverse Fourier transform of an integrable function $F(t)$
\[
f(x)=\mathscr{F}^{-1}[F](x)=\frac{1}{(2\pi)^n}\int_{\mathbb{R}^n}F(t)e^{ixt} dt
\]
and an integrable in $t$ function $F(t,y)$
\[
\mathscr{F}_t^{-1}[F](x,y)=\frac{1}{(2\pi)^n}\int_{\mathbb{R}^n}F(t,y)e^{ixt} dt.
\]
The definition of the Fourier transform of generalized functions of slow growth, see.\cite{Vla}.

      Consider the Dirichlet problem for the Tricomi-Keldysh type equation:
\[
y^m\Delta_xu+u_{yy}=0,~~x\in\mathbb{R}^n,~~ y>0,~~ m>-2,\eqno{(1.1)}
\]
\[
u(x,0)=\psi(x),~~x\in\mathbb{R}^n,\eqno{(1.2)}
\]
\[
u(x,y)~\text{is bounded as}~y\to\infty.\eqno{(1.3)}
\]

\section{Solution the Dirichlet problem for an equation of Tricomi type in the case
  $m=1$  by means the Fourier transform method}
\[
y\Delta_xu+u_{yy}=0,~~x\in\mathbb{R}^n,~~ y>0,\eqno{(2.1)}
\]
\[
u(x,0)=\psi(x),~~x\in\mathbb{R}^n,\eqno{(2.2)}
\]
\[
u(x,y)~\text{is bounded as}~y\to\infty.\eqno{(2.3)}
\]
Applying the Fourier transform  with respect to x to the equation (2.1) and denoting
\[
U(t,y)=\mathscr{F}_x[u](t,y),~~\Psi(t)=\mathscr{F}[\psi](t).
\]
we get the boundary value problem for ordinary differential equation with the parameter $t\in\mathbb{R}^n$:
\[
-y|t|^2U(t,y)+U_{yy}(t,y)=0,
\]
\[U(t,0)=\Psi(t),~~U(t,y)~\text{is bounded as}~y\to\infty.
\]
This is  Airy equation , its general solution is written in terms of the Airy functions
\[
U(t,y)=c_1(t)\textup{Ai}(|t|^{2/3}y)+c_2(t)\textup{Bi}(|t|^{2/3}y).
\]
Because  the
\[
\lim_{y\to\infty}\textup{Bi}(|t|^{2/3}y)=\infty~\text{and}~\textup{Ai}(0)=\frac{1}{3^{2/3}\Gamma(2/3)},
\]
then, taking into account the boundary conditions, we obtain the solution of the boundary problem
\[
U(t,y)=3^{2/3}\Gamma(2/3)\Psi(t)\textup{Ai}(|t|^{2/3}y).
\]
Applying the inverse Fourier transform, we obtain the solution of the original Dirichlet problem for the equation of Tricomi type (2.1) - (2.3) in the form of a convolution (if this convolution exists)
\[
u(x,y)=\psi(x)*k_{n1}(x,y),\eqno{(2.4)}
\]
where the kernel
\[
k_{n1}(x,y)=3^{2/3}\Gamma(2/3)\mathscr{F}_t^{-1}[\textup{Ai}(|t|^{2/3}y)](x,y).
\]
Let $|x|=r,~|t|=\rho,~\sigma_{n-1}$ - the area of the unit sphere in $\mathbb{R}^n$. To calculate the inverse Fourier transform we pass to spherical coordinates taking into account that for positive values of the argument the Airy function  is expressed via the Macdonald function
\[
\textup{Ai}(\rho^{2/3}y)=\frac{1}{\pi\sqrt{3}}\rho^{1/3}\sqrt{y}K_{1/3}(2/3y^{3/2}\rho),~~y>0.
\]
We have
\[
k_{n1}(x,y)=\frac{3^{2/3}\Gamma(2/3)}{(2\pi)^n}\int_{\mathbb{R}^n}\textup{Ai}(|t|^{2/3}y)e^{-ixt}dt=
\]
\[
=3^{2/3}\Gamma(2/3)\frac{\sigma_{n-1}}{(2\pi)^n}\int_0^{\infty}\textup{Ai}(\rho^{2/3}y)\rho^{n/2}d\rho\int_{0}^{\pi}e^{-ir\rho\cos \theta}\sin^{n-2} \theta d\theta=
\]
\[
=\frac{3^{2/3}\Gamma(2/3)}{(2\pi)^{n/2}r^{n/2-1}}\int_0^{\infty}\textup{Ai}(\rho^{2/3}y)\rho^{n/2}J_{n/2-1}(r\rho) d\rho=
\]
\[
=\frac{3^{2/3}\Gamma(2/3)\sqrt{y}}{(2\pi)^{n/2}r^{n/2-1}\pi\sqrt{3}}\int_0^{\infty}K_{1/3}(2/3y^{3/2}\rho)\rho^{1/3+n/2}J_{n/2-1}(r\rho) d\rho,
\]
where $J_{n/2-1}(r\rho)$~--~is Bessel function of the 1st kind of order $\nu=n/2-1$. This formula is valid for $n=1$, it is easy to check. The last integral is expressed in terms of the hypergeometric function $F$ (\cite{Rys} p. 684, the formula 6.576.3):
\[
\int_0^{\infty}K_{1/3}(2/3y^{3/2}\rho)\rho^{1/3+n/2}J_{n/2-1}(r\rho) d\rho=
\]
\[
=\frac{3^{n+1/3}\Gamma(n/2+1/3)}{2^{n/2+1}y^{3n/2+1/2}}F\left(\frac{n}{2}+\frac{1}{3},\frac{n}{2};\frac{n}{2};-\frac{9r^2}{4y^3}\right)=
\]
\[
=\frac{3^{n+1/3}\Gamma(n/2+1/3)}{2^{n/2+1}y^{3n/2+1/2}}\left(1+\frac{9r^2}{4y^3}\right)^{-n/2-1/3}.
\]
Finally, we obtain an expression for the kernel
\[
k_{n1}(x,y)=C_{n1}^*\frac{y}{(4y^3+9|x|^2)^{n/2+1/3}},~~x\in\mathbb{R}^n,~~y>0,
\]
\[
C_{n1}^*=\frac{3^{n+1/2}\Gamma(2/3)\Gamma(n/2+1/3)}{2^{1/3}\pi^{n/2+1}}.
\]
The received kernel has the following properties for  $y>0$:
\begin{align*}
1)~&k_{n1}(x,y)>0,\\
2)~&\int_{\mathbb{R}^n}k_{n1}(x,y)dx=1,\\
3)~&\forall\delta>0, \lim_{y\to+0} \sup_{|x|\ge\delta} k_{n1}(x,y)=0.
\end{align*}
     Property 1) is obvious. Property 2) follows from the fact that the Fourier transform of $k_{n1} (x,y)$ there are $3^{2/3}\Gamma(2/3)\textup{Ai}(|t|^{2/3}y)$,
\[
\int_{\mathbb{R}^n} k_{n1}(x,y)e^{ixt}dx=3^{2/3}\Gamma(2/3)\textup{Ai}(|t|^{2/3}y).
\]
Putting  $t=0$ , we obtain 
\[
\int_{\mathbb{R}^n} k_{n1}(x,y)dx=3^{2/3}\Gamma(2/3)\textup{Ai}(0)=1.
\]
 Property 3) follows from the fact that $k_{n1} (x,y)$ decreases monotonically as a function of $|x|$ .

       These properties mean that $k_{n1} (x,y)$ is \textit{the approximation to the identity} or $\delta$--shaped system of functions of $x$ (with parameter $y$ ), for $y\to+0 , k_{n1} (x,y)$  weakly converges to $\delta$--function $\delta(x)$.

      If  $\psi(x)$  is a bounded piecewise continuous function then convolution (2.4) exists and is recorded as an integral
\[
u(x,y)=C_{n1}^*\int_{\mathbb{R}^n}\frac{\psi(x)y}{(4y^3+9|x-t|^2)^{n/2+1/3}}dt.
\]
      From the fact that the kernel of integral is the approximation to the identity follows the equality
\[
\lim_{y\to+0} u(x,y)=\psi(x)
\]
at the points of continuity of $\psi(x)$, which means that the integral is a classical solution of the Dirichlet problem.
      For generalized functions of slow growth $\psi(x)\in\mathscr{S}'(\mathbb{R}^n)$  , for which a convolution exists, the function
\[
u(x,y)=\psi(x)*k_{n1}(x,y)
\]
is a generalized solution of the Dirichlet problem:
\[
\lim_{y\to+0} u(x,y)=\psi(x)~~  \text{in}~~  \mathscr{S}'.
\]
For example, if  $\psi(x)=\delta(x)$, then the solution of the Dirichlet problem is the kernel of the integral
\[
u(x,y)=\delta(x)*k_{n1}(x,y)=k_{n1}(x,y),
\]
\[
\lim_{y\to+0} k_{n1}(x,y)=\delta(x)~~\text{in}~~\mathscr{S}'.
\]
      If $\psi(x)\in L^p(\mathbb{R}^n),~1 \le p \le \infty$, then from the properties of the approximation to the identity [5] follows that
\[
\lim_{y\to+0} u(x,y)=\psi(x)
\]
for almost every $x$,  and if $p<\infty$ , then  $u(x,y)$ converges to  $\psi(x)$ in  the norm of $L^p(\mathbb{R}^n)$,  as $y\to+0$.

      For the case $n=1$ the integral, which gives the solution of the Dirichlet problem for the Tricomi equation (2.1) - (2.3), has the form
\[
u(x,y)=\frac{3^{3/2}\Gamma(2/3)\Gamma(5/6)}{\pi^{3/2}2^{1/3}}\int_{-\infty}^{\infty}\frac{y\psi(t)}{(4y^3+9(x-t)^2)^{5/6}}dt.
\]
      The kernel of the integral representation of the solution of the Dirichlet problem for multidimensional Tricomi equation, which is an approximation to the identity, represented in the form
\[
k_{n1}(x,y)=C_{n1}^*\frac{y}{(4y^3+9|x|^2)^{n/2+1/3}}=\frac{1}{y^{n3/2}}\varphi\left(\frac{|x|}{y^{3/2}}\right)
\]
where
\[
\varphi(r)=\frac{C_{n1}^*}{(4+9r^2)^{n/2+1/3}}.
\]
That is, the kernel  $k_{n1} (x,y)$  is the \textit{self-similar solution} of the Tricomi equation, which can be found by  the \textit{similarity method} (\cite{Bar},\cite{Pol}, Ch. 3). We use this method to solve  the Dirichlet problem for the equation of Tricomi-Keldysh type.

\section{Solution the Dirichlet problem for the equation of Tricomi-Keldysh type in the case  $m>-2$   by the similarity method}
We will search the kernel of the integral representation of the solution of the Dirichlet problem (1.1) - (1.3), which is an approximation to the identity, in the form of a self-similar solution of the equation of Tricomi-Keldysh type (1.1)
\[
u(x,y)=\frac{1}{y^{\alpha}}\varphi\left(\frac{r}{y^{\beta}}\right),~~\alpha>0,~~\beta>0,\eqno{(3.1)}
\]
where  $r=|x|$, that is, we are looking for a spherically symmetric solution, depending only on $|x|=r$. The equation of Tricomi-Keldysh type (1.1) for a spherically symmetric function takes the form
\[
y^m\left(u_{rr}+\frac{n-1}{r}u_r\right)+u_{yy}=0,~~y>0,~~m>-2,\eqno{(3.2)}
\]
To determine the constants $\alpha$ and $\beta$ we will do in equation (3.2) the change of variables
\[
u=C^l \bar u,~~r=C^k \bar r,~~y=C \bar y,~~(C>0),
\]
and require that this equation is transferred into itself. We obtain the equation in the new variables
\[
C^{m+l-2k}\left(\bar u_{\bar r\bar r}+\frac{n-1}{\bar r}\bar u_{\bar r}\right)+C^{l-2}\bar u_{\bar y\bar y}=0.
\]
In order to this equation coincides with the equation (3.2), we  set
\[m+l-2k=l-2.
\]
Hence
\[
k=\frac{m+2}{2} ,~~l~\text{is any}.
\]
In the new variables, the self-similar solution should have the same form (3.1)
\[
\bar u=\frac{1}{\bar y^{\alpha}}\varphi\left(\frac{\bar r}{\bar y^{\beta}}\right).
\]
Returning to the old variables, we obtain
\[
u=\frac{C^{\alpha+l}}{y^{\alpha}}\varphi\left(\frac{C^{-k+\beta}r}{y^{\beta}}\right),
\]
in order to coincide this  expression with the (3.1), we set
\[
\beta=k=\frac{m+2}{2},~\alpha=-l, \text{that is, any. We take}~ \alpha=kn=n\frac{m+2}{2}.
\]
 From the condition  $\alpha>0, \beta>0$  , it follows that $m>-2$. And so, we seek a solution of equation (3.2) in the form
\[
u=\frac{1}{y^{kn}}\varphi\left(\frac{r}{y^k}\right),~~k=\frac{m+2}{2},~~m>-2.\eqno{(3.3)}
\]
Substituting the function (3.3) in the equation (3.2), we obtain the equation
\[
(1+k^2r^2)\varphi''(r)+\left(\frac{n-1}{r}+2k^2nr+k^2r+kr\right)\varphi'(r)+kn(kn+1)\varphi(r)=0.
\]
By doing in this equation the change of variable
\[
1+k^2r^2=\xi,
\]
we obtain for the function $\bar \varphi(\xi)$ hypergeometric equation
\[
\xi(1-\xi)\bar \varphi''(\xi)+\left(\frac{n}{2}+1+\frac{1}{2k}-\xi\left(n+1+\frac{1}{2k}\right)\right)\bar \varphi'(\xi)-\left(\frac{n^2}{4}+\frac{n}{4k}\right)\bar \varphi(\xi)=0.
\]
We write it in the form
\[
\xi(1-\xi)\bar \varphi''(\xi)+\left(c-\xi(a+b+1)\right)\bar \varphi'(\xi)-ab\bar \varphi(\xi)=0,\eqno{(3.4)}
\]
where
\[
c=\frac{n}{2}+1+\frac{1}{2k},~~a=\frac{n}{2},~~b=\frac{n}{2}+\frac{1}{2k}.
\]
The general solution of the hypergeometric equation (3.4) has the form
\begin{align*}
\bar \varphi(\xi)=C_1F(a,b;c;\xi)+C_2\xi^{1-c}F(b-c+1,a-c+1;2-c;\xi)=\\
=C_1F\left(\frac{n}{2},\frac{n}{2}+\frac{1}{2k};\frac{n}{2}+1+\frac{1}{2k};\xi\right)+C_2\xi^{-n/2-1/2k}F\left(0,-\frac{1}{2k};1+\frac{n}{2}+\frac{1}{2k};\xi\right)=\\
=C_1F\left(\frac{n}{2},\frac{n}{2}+\frac{1}{m+2};\frac{n}{2}+1+\frac{1}{m+2};\xi\right)+C_2\xi^{-n/2-1/(m+2)},
\end{align*}
where $F(a,b;c;\xi)$ is the hypergeometric function. We will take a particular solution (denoting constant $C_2$   through  $C_{nm}$)
\[
\bar \varphi(\xi)=\frac{C_{nm}}{\xi^{n/2+1/(m+2)}}.
\]
Returning to the old variables, we obtain
\[
\varphi(r)=\frac{C_{nm}}{\left(1+\left(\frac{m+2}{2}\right)^2r^2\right)^{n/2+1/(m+2)}}.
\]
We choose a such constant $C_{nm}$   that the integral of $\varphi(|x|)$ over the entire space $\mathbb{R}^n$ is equal to unity. Passing to the spherical coordinates and denoting $\sigma_{n-1}$ area of the unit sphere in $\mathbb{R}^n$, we will have
\[
C_{nm}^{-1}=\int_{\mathbb{R}^n}\varphi(|x|)dx=\sigma_{n-1}\int_0^{\infty}\frac{r^{n-1}dr}{\left(1+\left(\frac{m+2}{2}\right)^2r^2\right)^{n/2+1/(m+2)}}=
\]
\[
=\frac{\sigma_{n-1}}{2\left(\frac{m+2}{2}\right)^n}\int_0^{\infty}\frac{t^{n/2-1}}{(1+t)^{n/2+1/(m+2)}}dt=\frac{\sigma_{n-1}2^{n-1}}{(m+2)^n}\frac{\Gamma\left(\frac{n}{2}\right)\Gamma\left(\frac{1}{m+2}\right)}{\Gamma\left(\frac{n}{2}+\frac{1}{m+2}\right)}.
\]
We made the change of variable $\left(\left(\frac{m+2}{2}\right)r\right)^2=t$, and used the formula
\[
\int_0^{\infty}\frac{t^{a-1}}{(1+t)^{b+c}}dt=\frac{\Gamma(a)\Gamma(b)}{\Gamma(a+b)},~a>0,b>0,
\]
where $\Gamma$ is Euler gamma function. Thus
\[
C_{nm}=\frac{(m+2)^n\Gamma\left(\frac{n}{2}+\frac{1}{m+2}\right)}{\pi^{n/2}2^n\Gamma\left(\frac{1}{m+2}\right)}.\eqno{(3.5)}
\]
Self-similar solution of the equation of Tricomi-Keldysh type, which is obtained by the formula (3.3), we denote
\[
k_{nm}(x,y)=\frac{1}{y^{n(m+2)/2}}\varphi\left(\frac{|x|}{y^{(m+2)/2}}\right)=
\]
\[
=\frac{C_{nm}y}{\left(y^{m+2}+\left(\frac{m+2}{2}\right)^2\right)^{n/2+1/(m+2)}},~~m>-2,\eqno{(3.6)}
\]
where $C_{nm}$ determined by the formula (3.5).

If  $m=0$, we obtain the Poisson kernel
\[
k_{n0}(x,y)=\frac{C_{n0}y}{(y^2+|x|^2)^{(n+1)/2}},
\]
where
\[
C_{n0}=\frac{\Gamma\left((n+1)/2\right)}{\pi^{(n+1)/2}}.
\]
      If $m=1$ we obtain the kernel of the integral representation of the solution of the Dirichlet problem (2.1) - (2.3), which was found in the previous section by Fourier transform
\[
k_{n1}(x,y)=\frac{C_{n1}y}{\left(y^3+\frac{9}{4}|x|^2\right)^{n/2+1/3}}=\frac{2^{n+2/3}C_{n1}y}{(4y^3+9|x|^2)^{n/2+1/3}},
\]
where
\[
2^{n+2/3}C_{n1}=\frac{2^{2/3}3^n\Gamma(n/2+1/3)}{\pi^{n/2}\Gamma(1/3)}=\frac{3^{n+1/2}\Gamma(2/3)\Gamma(n/2+1/3)}{2^{1/3}\pi^{n/2+1}}=C_{n1}^*.
\]
We show that the function $k_{nm} (x,y)$, defined by (3.6), is an approximation to the identity in the space of integrable on $\mathbb{R}^n$ functions, that is, has the following properties for $y>0$:
\begin{align*}
1)~&k_{nm}(x,y)>0,\\
2)~&\int_{\mathbb{R}^n}k_{nm}(x,y)dx=1,\\
3)~&\forall\delta>0, \lim_{y\to+0} \sup_{|x|\ge\delta} k_{nm}(x,y)=0.
\end{align*}
Properties 1) and 3) are proved like in section 2 for the case $m=1$. Let us prove property 2). We have
\[
\int_{\mathbb{R}^n}k_{nm}(x,y) dx=\int_{\mathbb{R}^n}\frac{1}{y^{n(m+2)/2}}\varphi\left(\frac{|x|}{y^{(m+2)/2}}\right)dx=\int_{\mathbb{R}^n}\varphi(|t|)dt=1.
\]
Therefore, the solution to the Dirichlet problem of the equation of Tricomi-Keldysh type can be written as a convolution of the boundary function $\psi(x)$ with the kernel $k_{nm} (x,y)$ (if a convolution exists)
\[
u(x,y)=\psi(x)*k_{nm}(x,y).
\]
If $\psi(x)$ is a bounded piecewise continuous function, the convolution exists and is recorded as an integral
\[
u(x,y)=C_{nm}\int_{\mathbb{R}^n}\frac{\psi(t)y}{(y^{m+2}+((m+2)/2)^2|x|^2)^{n/2+1/(m+2)}}dt.
\]
At points of continuity $\psi(x)$
\[
\lim_{y\to+0}u(x,y)=\psi(x).
\]
That is, the integral is the classical solution of the Dirichlet problem.

      For generalized functions of slow growth $\psi(x)\in\mathscr{S}'(\mathbb{R}^n)$ , for which a convolution exists, the function
\[
u(x,y)=\psi(x)*k_{nm}(x,y)
\]
is a generalized solution of the Dirichlet problem:
\[
\lim_{y\to+0} u(x,y)=\psi(x)~~  \text{in}~~  \mathscr{S}'.
\]
For example, if  $\psi(x)=\delta(x)$, then the solution of the Dirichlet problem is the kernel of the integral
\[
u(x,y)=\delta(x)*k_{nm}(x,y)=k_{nm}(x,y),
\]
\[
\lim_{y\to+0} k_{nm}(x,y)=\delta(x)~~\text{in}~~\mathscr{S}'.
\]
      If $\psi(x)\in L^p(\mathbb{R}^n),~1 \le p \le \infty$, then from the properties of the approximation to the identity [5] follows that
\[
\lim_{y\to+0} u(x,y)=\psi(x)
\]
for almost every $x$,  and if $p<\infty$ , then  $u(x,y)$ converges to  $\psi(x)$ in  the norm of $L^p(\mathbb{R}^n)$,  as $y\to+0$.

\textbf{Example.} For $n=1,m=-1$, we have the Dirichlet problem for the Keldysh  equation
\begin{gather*}
u_{xx}+yu_{yy}=0,~~-\infty<x<\infty,~~y>0,\\
u(x,0)=\psi(x),~~~-\infty<x<\infty,\\
u(x,y)~\text{is bounded as}~ y\to\infty.
\end{gather*}
If $\psi(x)$ is a bounded piecewise continuous function, the solution to this problem is given by the integral
\[
u(x,y)=2\int_{-\infty}^{\infty}\frac{y\psi(t)dt}{(4y+(x-t)^2)^{3/2}}.
\]
Take $\psi(x)=\left\{\begin{aligned}a,~x<0\\b,~x>0\end{aligned}\right.$, then
\[
u(x,y)=2ay\int_{-\infty}^0\frac{dt}{(4y+(x-t)^2)^{3/2}}+2by\int_0^{\infty}\frac{dt}{(4y+(x-t)^2)^{3/2}}=
\]
\[
=\frac{a+b}{2}+\frac{b-a}{2}\frac{x}{\sqrt{4y+x^2}}.\eqno{(3.7)}
\]
It is easy to verify that the function (3.7)
\begin{enumerate}
\item
satisfies the Keldysh equation~$u_{xx}+yu_{yy}=0,~y>0$,
\item
is bounded,~$|u(x,y)|\le\max (|a|,|b|),~y>0$,
\item
satisfies the boundary condition,
\[
\lim_{y\to+0}u(x,y)=\frac{a+b}{2}+\frac{b-a}{2}\frac{x}{|x|}=\left\{\begin{aligned}a,~x<0\\b,~x>0\end{aligned}\right.=\psi(x).
\]
\end{enumerate}
At the point of discontinuity 
\[
\lim_{y\to+0}u(0,y)=\frac{a+b}{2}.
\]

\section*{                                             Conclusion}
    In the paper is considered an elliptic equation in a multidimensional half-space
\[
y^m\Delta_x+u_{yy}=0,~~x\in\mathbb{R}^n,~~ y>0,~~ m>-2,\eqno{(\textup{T})}
\]
which is a generalization of the equations of Tricomi ($n=1,m=1$), Gellerstedt ($n=1,m>0$), Keldysh ($n = 1,-2 <m <0$) and Laplace ($n\ge1,m = 0$).
We are looking for the solution  of this equation, which is bounded when $y\to\infty$ and satisfies on the boundary of the half-space Dirichlet boundary condition
\[
u(x,0)=\psi(x),~~x\in\mathbb{R}^n.
\]
Is found an exact solution of the Dirichlet problem in the form of an integral that generalizes the well-known Poisson formula giving the solution of the Dirichlet problem for the Laplace equation. The kernel of the integral representation of the solution of the Dirichlet problem for the case $m = 1$ is found by the Fourier transform method, and for the general case of $m> -2$ by the similarity method in form a self-similar solution of equation (T). It is shown that this kernel is an approximation to the identity. The general properties of the approximation to the identity implies that the integral is a classical solution of the Dirichlet problem if the boundary function is bounded and piecewise continuous, and if the boundary function is a generalized function of slow growth, and exists a convolution of the function with the kernel, the convolution is a generalized solution of the Dirichlet problem . That is, a solution for $y\to + 0$ tends to the boundary function in $\mathscr{S}'(\mathbb{R}^n)$ in the sense of theory of generalized functions .

\end{document}